# Electrostatic PIC with adaptive Cartesian mesh

**Vladimir Kolobov**[1,2] **and Robert Arslanbekov**[1]

[1] CFD Research Corporation, Huntsville, AL, USA
[2] The University of Alabama in Huntsville, Huntsville, AL 35899, USA

E-mail: vladimir.kolobov@cfdrc.com

**Abstract**. We describe an initial implementation of an electrostatic Particle-in-Cell (ES-PIC) module with adaptive Cartesian mesh in our Unified Flow Solver framework. Challenges of PIC method with cell-based adaptive mesh refinement (AMR) are related to a decrease of the particle-per-cell number in the refined cells with a corresponding increase of the numerical noise. The developed ES-PIC solver is validated for capacitively coupled plasma, its AMR capabilities are demonstrated for simulations of streamer development during high-pressure gas breakdown. It is shown that cell-based AMR provides a convenient particle management algorithm for exponential multiplications of electrons and ions in the ionization events.

1. Introduction

Particle-in-Cell (PIC) is a mature technology widely used for plasma simulations. There are three versions of PIC: an electrostatic (EC) PIC uses Poisson equation for calculating electric field, an electromagnetic (EM) PIC employs full-wave Maxwell solver for calculations of electric and magnetic fields, and a quasi-static (QS) PIC uses a quasi-static version of the Maxwell equations (such as Darwin model) for calculation of the EM fields. Surprisingly, PIC solvers with adaptive mesh refinement (AMR) are relatively rare. It is worth mentioning the WARP code for accelerator applications [1], and an explicit EM-PIC code of Fujimoto [2] for astrophysics applications, both using block-structured AMR.

In this paper, we describe an initial implementation of an ES-PIC module in our Unified Flow Solver (UFS) framework [3]. UFS is designed for hybrid simulations of partially-ionized collisional plasmas using adaptive Cartesian mesh and automatic selection of kinetic and fluid solvers for transport processes. This is called Adaptive Mesh *and* Algorithm Refinement (AMAR) methodology. A cell-by-cell selection of most suitable solvers for different plasma species is based on continuum-breakdown criteria that can be specified in advance by the user.

The PIC module in UFS supplements already developed fluid models, mesh-based kinetic solvers (Boltzmann, Vlasov, Fokker-Plank) and the particle-based Direct Simulation Monte Carlo (DSMC) and Photon Monte Carlo (PMC) solvers previously developed for simulations of mixed rarefied-continuum flows, radiation transport and plasma dynamics.

Compared to neutral gases, plasma simulations pose extra challenges due to the disparity of time and spatial scales associated with small electron mass [4]. Criteria for selecting kinetic and fluid models for electrons, ions and neutrals are different, and strongly depend on plasma conditions. In collisionless magnetized plasmas, ions are often treated kinetically and electrons are assumed to be a fluid. In gas discharge physics, electrons are commonly treated kinetically, and fluid models applied for ions and neutrals [5]. First steps towards adaptive kinetic-fluid plasma simulations have been made. Two-way coupling of a global Hall magnetohydrodynamics with a local implicit PIC model (MHD with Embedded PIC regions (MHD-EPIC)) has been recently demonstrated for space plasmas.[6]

---
[1] To whom any correspondence should be addressed.

In the present paper, we first provide some details of the UFS methodology and discuss parallel strategy for the AMAR codes. Then, we describe an initial implementation of the ES-PIC module in UFS focusing on challenges associated with AMR. Finally, validation of the ES-PIC solver for capacitively coupled plasma (CCP) and demonstration of its AMR capabilities for simulations of streamer breakdown are described. We show that cell-based AMR technique provides a convenient adaptive particle management algorithm for exponential multiplication of charged particles in PIC codes.

## 2. Unified Flow Solver

Basic architecture of UFS is shown in Figure 1. The AMAR core is implemented on top of Gerris Flow Solver (GFS) - an open source computing environment for solving partial differential equations with AMR [7]. GFS provides automatic mesh generation for complex geometries, portable parallel support using the MPI library, dynamic load balancing, and parallel offline visualization. GFS physics includes time-dependent incompressible variable-density Euler, Stokes or Navier-Stokes equations with volume of fluid method for interfacial flows. A coarse-grained parallelization algorithm is based on the "Forest of Trees" methodology [8].

UFS enables the higher degree of adaptation by using different physical models in different parts of computational domain. In particular, the computational domain could be decomposed into kinetic and fluid cells using physics-based continuum breakdown criteria. This methodology was first implemented for mixed rarefied-continuum flows and later extended for hybrid simulations of radiation transport coupling a Photon Monte Carlo (PMC) solver with a diffusion model of radiation transport selected based on the local photon mean free path.

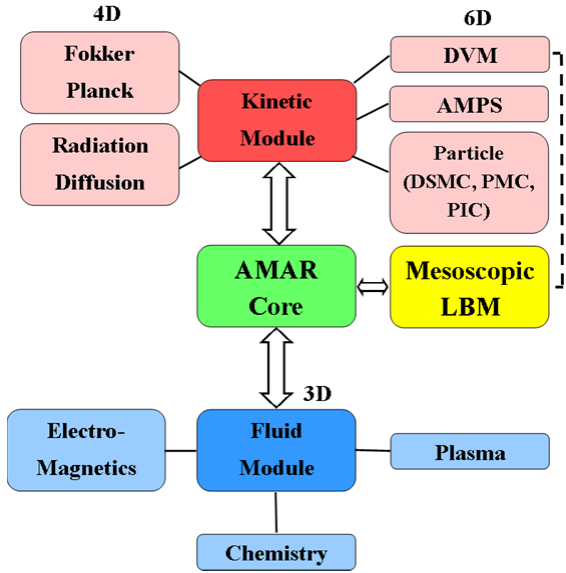

Figure 1: UFS Architecture.

The Kinetic Module in UFS can solve Boltzmann, Vlasov, and Fokker-Planck kinetic equations using Discrete Velocity Method (DVM). The recently developed Adaptive Mesh in Phase Space (AMPS) methodology [9] can adapt mesh in both physical and velocity spaces. In addition, the Kinetic Module has the particle-based DSMC [10] and PMC solvers [11]. The mesoscopic Lattice Boltzmann Method (LBM) uses a minimal set of discrete velocities as a subset of the DVM kinetic solvers [12].

The Fluid Module in UFS contains Euler and Navier-Stokes solvers for reacting gas mixtures using different numerical schemes. Flow solvers with adaptive Cartesian mesh offer a viable alternative to the conventional CFD methods using body-conforming mesh. The Immersed Boundary Method helps improving calculations of skin friction and heat fluxes for supersonic viscous flows [13].

For charged species (electrons and ions), multi-temperature drift-diffusion models are used, which are coupled to the Poisson solver for the electrostatic field [4].

*2.1. Parallel Algorithms for Accelerated Heterogeneous Computing*

Parallel capability is a must of today's codes. Modern systems often use GPU-accelerators for massively parallel tasks. Porting adaptive kinetic-fluid solvers to heterogeneous accelerator-based systems poses several challenges associated with irregular data structures of AMR, vastly different costs of computing in fluid and kinetic cells, and dynamic balancing of CPU and GPU loads during grid adaptations and model refinement.

UFS uses two methods of domain decomposition for parallel computing: a fine-grained method using Space Filling Curves (SFC) and a coarse-grained using Forest of Trees (FOT). The SFC algorithm is currently based on Morton order. After ordering of cells, a weight is assigned to each cell based on the node time required to perform computations in this cell. Kinetic cells have a much larger weight compared to the fluid cells. Furthermore, the array of cells modified with corresponding weights, is subdivided into sub-arrays equal to the number of nodes, in such a way that the weights of each sub-arrays are approximately the same. This method allows very precise and efficient domain decomposition and dynamic load balancing between different nodes.

The FOT algorithm consists of dividing the computational domain into boxes of fixed size with a Cartesian tree built in each box. Each tree becomes the smallest parallel subdomain. The trees are connected through their common boundaries. Graph partitioning algorithms are used for domain decomposition and Dynamic Load Balancing (DLB). This represents a coarse-grain parallelization algorithm since the CPU load balancing can be done only in terms of the building (root) boxes. Each node operates on a given number of boxes, and the computational grid is stored only on the node it belongs to. During grid adaptation these boxes are exchanged between the nodes for DLB. Each box has its own ID number and a set of boxes on each host node share the same Processor or node ID (PID) number. The box-based DLB has the benefit of efficient memory allocation because only the trees that belong to a node need to be kept in the node's memory.

It is important to emphasize key differences between computational load of cells in particle-based and grid-based kinetic solvers. In the particle-based solvers, the number of particles per cell should be kept approximately constant and the cell size should be about or smaller than the local mean free path or Debye length (for explicit PIC codes). Thus, the mesh adaptation for particle-based solvers does not increase the computational load if the particle density remains approximately the same throughout the computational domain. The computational load of mesh-based kinetic solvers is directly proportional to the number of cells in physical space. The weight of each kinetic cell (in physical space) is proportional to the number of cells in velocity space. Fluid cells have small weight compared to the kinetic cells. Adaptation of spatial mesh changes the cost of computations directly proportional to the number of cells in physical space with corresponding weights for the kinetic and fluid cells. These differences are of principal importance for the development of parallel algorithms for adaptive kinetic-fluid solvers based on particle- and mesh-based schemes.

So far, we have used the simplest approach where GPUs were applied for the kinetic cells whereas the fluid cells were computed on CPUs. This methodology has been recently demonstrated for simulations of mixed rarefied-continuum flow over a hypersonic scramjet on a CPU-GPU cluster using DVM Boltzmann and Navier-Stokes solvers [12].

### 3. An electrostatic Particle-in-Cell Module in UFS

We have implemented an electrostatic PIC module in UFS with a Monte Carlo Collision (MCC) algorithm describing collisions of electrons and ions with neutrals. Our multi-grid Poisson solver calculates the electric potential at cell centers. We first implemented the particle-field coupling via the nearest grid point (NGP) method to assign the particle space charges at cell centers. For calculation of the electric field at particle location, $\vec{E}_n(\vec{x}_n)$, we used bi-linear interpolation based on components of the electric field vector on cell nodes/corners. The AMR capabilities of the PIC module using the first-order accurate NGP technique are demonstrated below.

Both explicit and implicit PIC schemes have been implemented. Explicit PIC requires grid spacing to be smaller than ~3 Debye lengths ($\Delta x < 3\lambda_D$) and the time step to be smaller than the inverse electron plasma frequency. Implicit PIC can handle larger cell sizes and time steps. Larger cell sizes also allow using smaller numbers of particles while keeping acceptable particle per cell (PPC) parameter. We have chosen the direct implicit method of DIPIC [14]. In order to implement the DIPIC scheme in UFS, we have extended the Poisson solver with the implicit susceptibility. The implicit scheme allowed simulating higher density plasma with smaller number of particles (see below).

Having implemented and tested the NGP scheme, we used the more accurate cloud-in-cell (CIC) approximation to reduce noise (improve statistics). Both the momentum conserving scheme (CIC-EM) and energy-conserving scheme (CIC-EC) [15] have been implemented for uniform mesh. Challenges of CIC implementation on adaptive Cartesian mesh are discussed below.

### 3.1. Adaptive Cartesian Mesh

In terms of particle handling on the AMR grid, our PIC code follows the DSMC code previously described in Ref. [10]. A single grid is used for particle movement, collisions and visualization. The quad/octree data structure allows straightforward and efficient data management during dynamic grid refinement/coarsening and makes possible seamless parallelization of the code. However, particle movement/tracking becomes more involved compared to a uniform Cartesian grid. Particles are assigned to each (leaf) cell using cell-based pointers and the global particle-list is used only for easy bookkeeping purposes. The particle data storage in UFS-PIC is thus cell-based, which means that each cell has a list of particles assigned to it. When a particle crosses a face of a cell, it is removed from the list of its old cell and added to the list of a neighbour cell. Particles are moved inside each computational cell according to Newton's laws until they hit a cell face or a solid panel face. Particle tracing is efficient since all full cells are square or cubic. In cut-cells at non-conforming solid boundaries, the solid panel face is checked for possible reflection, and appropriate boundary conditions are applied when this face is crossed. Since a cell neighbour can be at a higher level, a new particle location is assigned depending on which part of the face has been hit by the particle.

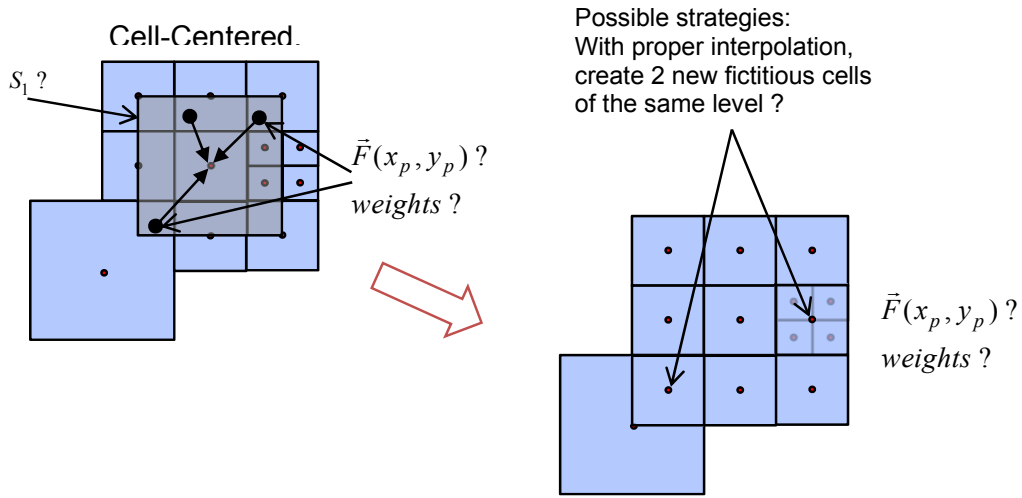

Figure 2. Illustration of the CIC method for a uniform grid (left) and adaptive grid (right). $S_1$ is the shape function used in the scatter step and $\vec{F}(x_p, y_p)$ function is used in the gather step.

No special treatment (such as assigning different weights) is carried out to account for different volumes of cells on the adapted grid. Moreover, no special treatment of small cut cells is performed.

The latter can lead to significant scattering of the surface data. To tackle this issue, a special routine is implemented which traverses all cut cells at a given coarse level (usually corresponding to initial uniform grid). Inside each such cell, all leaf cut cells are visited and the surface data are weight averaged with solid panel surface area. This procedure significantly reduces statistical scatter in the calculated surface fluxes, and provides similar advantages to a decoupling of surface elements from the grid structure. During refinement of a coarse cell, particles are stored in new cells according to their position in this coarse cell. During cell coarsening, particles in fine cells are assigned to a new coarse cell. Statistics counters are reset to zero when the AMR grid changes.

The implementation of CIC for cell-based adaptive Cartesian mesh is a subject of current research To derive the interpolation functions for the CIC approach with momentum and energy conservation constrains, as well as variable particle weights when particles travels between cells of different refinement levels, we have considered some possible strategies to application of the existing technologies to the adapted grids, which consists in creating fictitious (ghost) cells (see Figure 2) to mimic a uniform grid behavior with some proper field and particle interpolations. Such cells can be created in neighbor (both direct and diagonal) cells as finer (+1) and coarser (−1) level cells.

The charge conservation method requires using the same shape function during the scatter and gather steps. For a uniform grid, the shape function covers the region shown in Figure 2, left. For AMR, we use a similar shape function, which involves direct and diagonal neighbor cells of different levels of refinement (see Figure 2, right). Implementation of a better interpolation technique developed in [16] in our PIC-AMR code is expected to provide better accuracy. This technique allows a linear function defined on cell centers of the AMR grid to be ideally restored for any neighbor cell configuration (location and cell level).

### 3.2. MPI parallelization of the PIC module using FOT algorithm

The MPI parallelization of the PIC module follows closely that of the DSMC module. The FOT methodology is used with extra data exchange between space charges and acceleration vectors. Thus, good scaling and processor load have been observed using PIC with increasing number of processors for close-to-uniform spatial distributions of charged particles. For highly non-uniform spatial distributions of the particles and variable PPC number, implementing good DLB requires new graph partitioning algorithm taking into account different weight of cells (which is proportional to PPC number). It is expected that the FOT algorithm is preferable to SFC for the PIC code, because of multigrid method used by the Poisson solver.

In future work, we plan to implement DLB capabilities for particle-based kinetic solvers, which are required for adaptive grids and varying PPC number. The DLB module will use the box-exchange method and the graph partitioning algorithm currently available for the fluid solvers and is now solely based on the number of cells per partition. We will extend this algorithm for cells with different weights by implementing additional criteria for the graph partitioning algorithm.

### 3.3. Validation of the ES-PIC solver for CCP discharges

We have validated the UFS-PIC module for a Capacitively Coupled Plasma (CCP) benchmark described in a topical review [17]. This review discusses several issues of the PIC codes (sufficient representation of energetic particles, statistical noise, and numerical heating). The benchmark is a CCP discharge in Ar gas pressures 50 and 100 mTorr between two planar electrodes separated by a 3 cm gap driven by an RF voltage at a frequency of 15.36 MHz and voltage amplitude of 100 V.

The fidelity of PIC-MCC simulations strongly depends on the number of super-particles. The number of super-particles in our simulations for converged results was found to be ~150K for 50 mTorr and ~25–50K for 100 mTorr, which makes the spatially average PPC parameter of ~2,000 and ~300–600, respectively.

Figure 3 shows results of UFS-PIC simulations in terms of RF-cycle averaged spatial distributions of electron and ion densities, electron temperature and electrostatic potential, as well the electron energy distribution function (EEDF). Also shown are data from Ref. [17] for direct comparison. One can see that the agreement for all plasma parameters (electron and ion densities, electron temperatures and electrostatic potential) is very good for both pressures of 50 and 100 mTorr, in spite of the differences in the included electrode processes in these simulations. In our simulations, no electron emission from electrodes was assumed. The simulations in Ref. [17] have taken into account secondary electron emission from electrodes due to ion bombardment with a coefficient of 0.2. The latter can lead to the presence of energetic electrons in the space-charge sheath and to slightly different profiles of plasma densities and especially electron temperature in the sheaths. One can also observe that the electron temperature profiles at 50 and 100 mTorr are drastically different, which can be explained by the CCP operational mode transition taking place between 50 and 100 mTorr. The predicted RF-cycle averaged EEDFs are in excellent agreement with those obtained in Kim et al.[17].

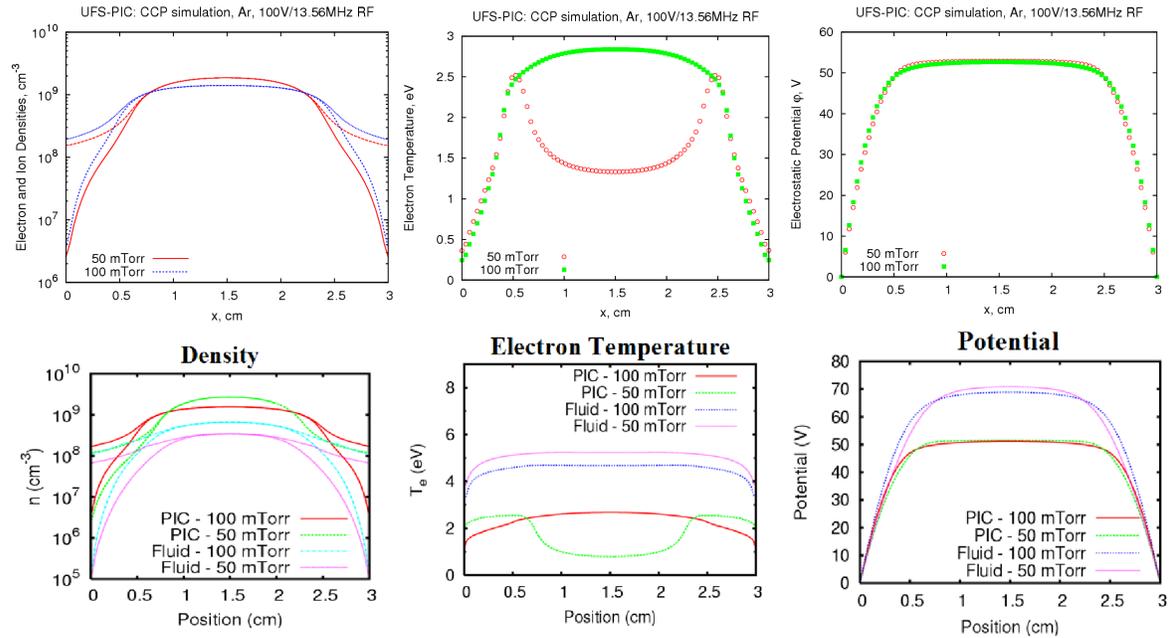

Figure 3. Comparison UFS-PIC results (top) with Kim et al.[17] results (bottom) for a benchmark CCP problem: RF-cycle averaged electron and ion densities (left), temperatures (center) and electric potential (right).

The accuracy of a PIC-MCC simulation strongly depends on the number of super-particles [17]. The best agreement with experimental data in Kim et al. [17], was obtained when using a sufficiently large number of super-particles of ~150K. This number of super-particles corresponds very well to our findings in the conducted convergence studies. Recall that we have also used ~150K super-particles for our 50 mTorr CCP simulations presented above. In Figure 4, we compare the UFS-PIC results obtained at different (spatially-average) PPC number of 2,000 (total number of particles ~150K), 750 and 350. One can see that only when (spatially average) PPC number reaches ~2,000 do the results converge and agree with the data in Kim et al. [17].

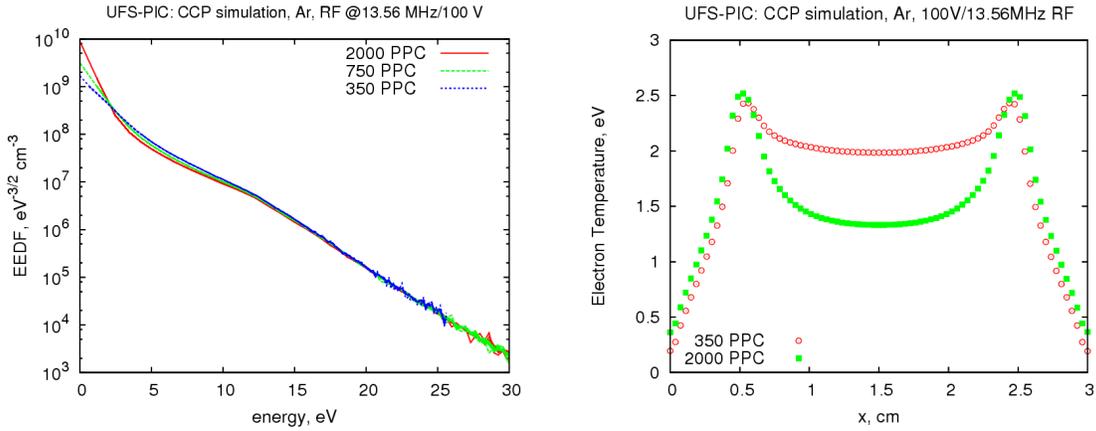

Figure 4. UFS-PIC simulation of CCP with different PPC parameters: RF-cycle averaged EEDF at plasma centre (left) and profiles of electron temperature (right).

The fact that PPC parameters of the order of a thousand are required provides a clear proof for the necessity of a hybrid approach for multi-dimensional simulations of realistic systems. Simulations of CCP discharges by a PIC method with such a large PPC make any realistic 3D and even 2D simulation impractical. It is expected that the low-energy electrons, controlling slow plasma processes (e.g., ambipolar diffusion, plasma densities, etc), can be described by alternative models (such as fluid models or Fokker-Planck kinetic solvers) while the high-energy electrons responsible for ionization can be described by the PIC methods. For a Maxwellian distribution with a temperature $T_e$, the ratio of high- to low energy electrons is proportional to $\exp(-\varepsilon_i / T_e)$, where $\varepsilon_i$ is the ionization threshold.

## 4. Simulations of High-Pressure Gas Breakdown

We illustrate below 3D simulations of gas breakdown in Ar at 100 Torr with inter-electrode distance of 40 mm inspired by the experimental studies of high-pressure gas breakdown between a needle-like cathode and a planar anode [18] shown in Figure 5. It is seen how an ionization wave propagating from a pin cathode reaches the anode after ~10 ns, and then returns to the cathode after ~25 ns. We were able to simulate the initial stage of this process with our PIC and fluid models.

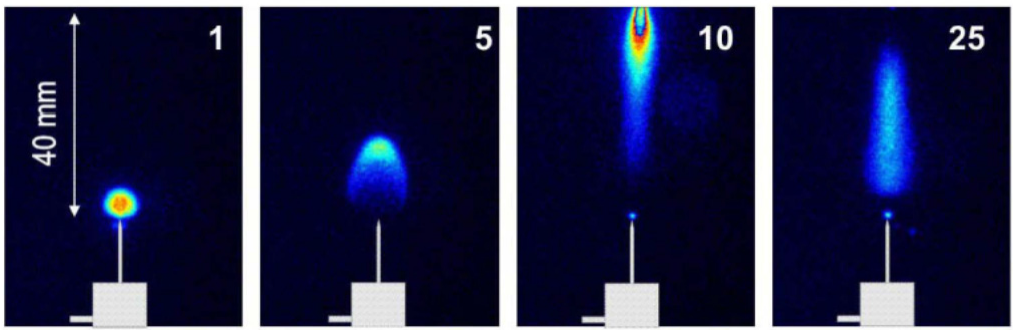

Figure 5. ICCD images of nanosecond discharge developing in air initiated by 400-ns HV pulse with short (11 ns) rise time. Numbers correspond to camera gate delay (in nanoseconds) after the pulse appearance on the HV electrode. Air, 680 Torr (adapted from Ref. [18]).

*4.1. PIC simulations with AMR*

Figure 6 shows results of 3D PIC simulations for applied voltages of 3.25 and 5 kV with the numbers of super-particles about 8M and 15M, respectively. At the high applied voltage, the breakdown occurs faster (~35 ns vs ~55 ns) and higher electron densities are attained, as expected. At the lower voltage, there is almost no branching, whereas quite significant branching occurs at the higher applied voltage. Similar branching was reported in simulations with uniform Cartesian mesh [19]. The AMR capabilities of the PIC are crucial for efficient simulations of the breakdown dynamics. However, even with AMR, we were not able to simulate breakdown at pressures above 100 Torr, and could not simulate the return strike observed in the experiments due to the very high computational cost of the PIC method under these conditions.

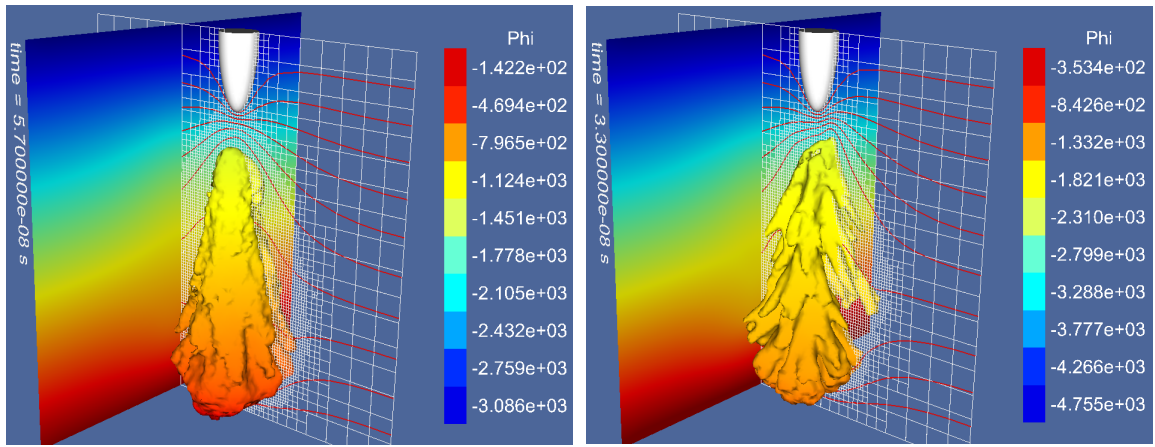

Figure 6. UFS-PIC simulation of breakdown for applied voltage of 3.25 kV (left) and 5 kV (right). Shown are isosurfaces of electron density (at $10^{16}$ m$^{-3}$ and $5\times10^{16}$ m$^{-3}$ levels), adapted grids, and isolines of electrostatic potential at 57 (left) and 33 (right) ns.

*4.2. Fluid Simulations*

In addition to the PIC simulations described above, we have simulated the same problem using the fluid plasma model. The fluid model results shown in Figure 7 are quite different compared to the PIC results in Figure 6.

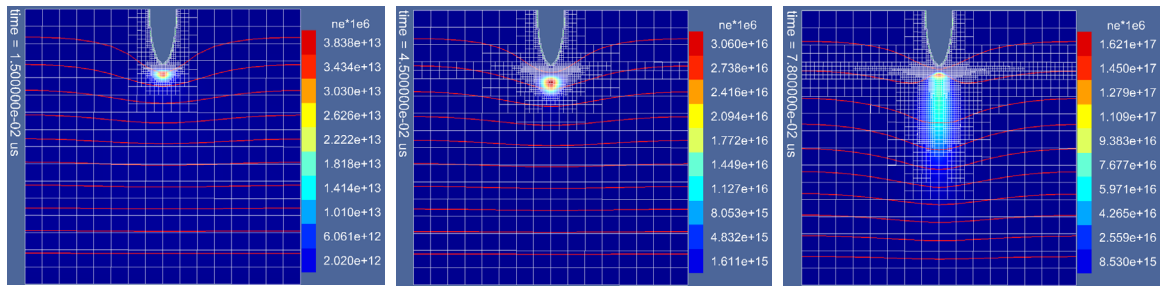

Figure 7. Fluid simulation of plasma breakdown in 100 Torr Ar gas with inter-electrode gap of 40 mm. Shown are profiles are electron density, adapted grids, and isolines of electrostatic potential at 3 time instances of 15, 45 and 78 ns.

In particular, the plasma channel predicted by the fluid model is narrower, and the time when the ionization wave reaches the anode surface (~90 ns) is longer than that predicted by the PIC. At 100

Torr, kinetic effects (such as non-local ionization by fast electrons, electron energy transport, etc.) are expected to play an important role in the breakdown process. Kinetic effects should also play a role at higher pressures in the high-field regions near the streamer front where the high-energy electrons are produced. Thus, hybrid PIC-fluid models are highly desirable for simulations streamer-spark-arc discharges. Using PIC only for the fast electrons in small areas near the steamer front and fluid model elsewhere can drastically reduce the computational cost without sacrificing the accuracy of simulations [19].

*4.3 Discussion*
One of the main problems in developing PIC-AMR codes is related to a decrease of the PPC number in the refined regions with a corresponding increase of the numerical noise.[20] The remedy for this problem may be the particle splitting-coalescence processes.[21] These processes should conserve such quantities as the charge and current densities, the total charge and mass of particles, the total momentum and energy of particles, and the distribution function of particles. For example, in the splitting algorithm in a 2D system, a particle is replaced by four particles. The particle merging algorithm has been refined and improved in Ref [22] using a k-d tree method for searching the nearest neighbors. Having efficient quad/octree generation and management (such as fast traversal/search) algorithms in the UFS framework, it would be relatively straightforward to implement such algorithms in our code. However, our particle handling on the AMR grid (see Section 3.1) could make this unnecessary.

Recently, a particle merging algorithm for problems where the number of particles grows exponentially in time has been proposed [23]. Such conditions commonly occur during gas breakdown. In our simulations of high-pressure streamer breakdown, the number of charged particles quickly increased from a few thousands to 10-20M during the breakdown process. Using dynamic AMR, the PPC parameter was kept at the same level (~500-1000). As often happens in plasma systems, refined cells are located in regions with small Debye length where the plasma density is large (AMR helps resolving the Debye length for PIC). Therefore, in most cases, the refined cells have sufficient number of particles per cell, and the particle splitting is not needed. In these cases, the AMR technique provides a convenient adaptive particle management algorithm.

In future work, we plan to use a hybrid solver where PIC model is used only in selected areas of computational domain whereas fluid model is used in other parts of the domain. Criteria for selecting kinetic patches could be based on the ratio of characteristic plasma scale (based on local gradients of plasma density) and the energy relaxation length for electrons (taking into electron collisions with atoms and the Coulomb interactions among charges particles).

## 5. Conclusions
We have described an initial implementation of an electrostatic Particle-in-Cell module with adaptive Cartesian mesh in our Unified Flow Solver framework. ES-PIC-AMR has been validated for a capacitively coupled plasma benchmark. Its AMR capabilities have been demonstrated for simulations of streamer development during high-pressure gas breakdown. The cell-based AMR technique provided a convenient adaptive particle management algorithm for problems with exponential multiplication of particles.

With the addition of the PIC module, we have now a full set of fluid and kinetic solvers (both grid-based and particle-based) that can be applied for hybrid plasma simulations. Future work will be devoted to identifying physics-based criteria for selecting appropriate kinetic and fluid solvers for different plasma components in different parts of the computational domain for applications requiring adaptive kinetic-fluid solvers.

**Acknowledgments**
This work was partially supported by the DARPA SBIR Project W31P4Q-15-C-0047 and by the US Department of Energy Office of Fusion Energy Science Contract DE-SC0001939.